# Assessing the influence of social media feedback on travelers' future trip-planning behavior: A multi-model machine learning approach

Sayantan Mukherjee[a], Pritam Ranjan[b*], Joysankar Bhattacharya[c]

[a]Communications Area, Indian Institute of Management Indore, MP, India
[b]Operations Management & Quantitative Techniques Area, Indian Institute of Management Indore, MP, India
[c]Economics Area, Indian Institute of Management Indore, MP, India

*Corresponding author email:* [pritamr@iimidr.ac.in](pritamr@iimidr.ac.in)

**Abstract:** With the surge of domestic tourism in India and the influence of social media on young tourists, this paper aims to address the research question on how "social return" – responses received on social media sharing – of recent trip details can influence decision-making for short-term future travels. The paper develops a multi-model framework to build a predictive machine learning model that establishes a relationship between a traveler's social return, various social media usage, trip-related factors, and her future trip-planning behavior. The primary data was collected via a survey from Indian tourists. After data cleaning, the imbalance in the data was addressed using a robust oversampling method, and the reliability of the predictive model was ensured by applying a Monte Carlo cross-validation technique. The results suggest at least 75% overall accuracy in predicting the influence of social return on changing the future trip plan. Moreover, the model fit results provide crucial practical implications for the domestic tourism sector in India with future research directions concerning social media, destination marketing, smart tourism, heritage tourism, etc.

## 1. Introduction

Tourism and travel have long been associated with individuals using them to strengthen their social ties and social status. Travel thus becomes both an identity marker and a core part of self-construction (McCabe & Stokoe, 2010), and tourists' intentions to share positive destination content often stem from how they perceive their experience (Luna-Cortés, 2021). Social media (SM) has deepened this connection between travel and social standing. The immediacy and visual richness of social media platforms often inspire aspirational travel, encouraging users to emulate the experiences of others (Munar & Jacobsen, 2014). Furthermore, the authenticity and relatability of shared content tend to carry more persuasive weight than traditional promotional materials. As a result, tourism patterns are increasingly driven by digital interactions (Shen & Wall, 2020), such as SM platforms.

Social networking sites facilitate interactions with family, friends, colleagues, neighbors, and even strangers (Boto-García et al., 2022), especially among individuals from Generations Y and Z (Mude & Undale, 2023). According to the Global Social Media Statistics (Kepios, 2025), there were an estimated 5.3 billion SM users in 2025, representing a 4.7 % increase year-on-year. The market size for SM is expected to reach more than $286 billion in 2025, a 13.3% increase from 2024 (Chaffey, 2025). In response, destination marketing organizations actively use SM dynamics to influence tourist preferences (Morosan, 2013). Our work investigates the relationship between the availability of self-generated SM content and its influence on future trip-planning behavior for interconnected and active individuals on SM.

The positive response/feedback on one's (past) SM content can be labeled as the "social return" (SR) (Boley et al., 2018; 2023). The concept of "return" from travel extends beyond financial or material gain, encompassing emotional, experiential, and social dimensions. Travelers often seek personal enrichment, cultural understanding, and emotional well-being as key outcomes of their journeys. Moreover, shared travel narratives contribute to collective knowledge and influence the travel aspirations and decisions of others within one's network. Although some studies have examined the notion of "return" (Pocock & McIntosh, 2011) and SR (Boley et al., 2018, 2023) in long-term travel experiences marked by emotional transformation, little is known about SR in the context of short-term travel planning influenced by SM sharing.

We further consider the concept of anticipated social return (ASR) when potential travelers consider SR as an influencing factor in their future trip planning behavior. Therefore, we need to investigate whether ASR is present or not in a traveler's trip planning discourse. Based on our current information, no known studies have investigated ASR using a data-driven, predictive modeling approach grounded in machine learning (ML) techniques in the Indian context—a rapidly growing and digitally active tourism market. Our study addresses these gaps by shifting the analytical lens toward social media–oriented predictors of trip-planning behavior, such as SM connections, sharing attitudes, content types (e.g., photos taken/shared), and engagement metrics (likes, comments, etc.) for Indian travelers of generations Y and Z.

India has witnessed a remarkable surge in domestic tourism, with a 155.48% increase in domestic tourist visits in 2022 (India Tourism, 2023). Concurrently, India ranks highest in Facebook user base globally, with 383.5 million users (Statista, 2025a), and millennials and Gen Z constitute the majority—88.6% on Facebook and 93% on Instagram (NapoleonCat, 2025a, 2025b). Moreover, the Gen Z population is called digital natives (Chang et al., 2023), and their purchase decisions, including travel, are heavily influenced by SM (Pan & Satchabut, 2022). Additionally, the young generation is prone to digital fatigue (Dave et al., 2024) and fear of missing out (FOMO) (Harahap et al., 2024) when it comes to searching for information online and making decisions. Despite a strong connection between SM engagement and travel behavior, limited research addresses how young Indian travelers use self-generated SM content and the reactions it garners from others (i.e., SR) for future trip planning (e.g., Mittal et al., 2022). The scarcity of research addressing this specific aspect becomes another motivation for our research endeavor.

Therefore, by advancing a novel conceptualization of ASR and employing a machine learning–enhanced methodology, this study not only fills a critical gap in tourism scholarship but also opens up new directions for understanding how SM engagement drives travel intentions in digital-first markets. As such, it lays the groundwork for a more granular, predictive, and socially embedded theory of trip-planning behavior.

*Research Objective:* To investigate "Whether feedback on previous travel experiences related posts on SM, along with other aspects of a traveler's SM usage behavior, impact the participant's future trip planning?". To carry out this exercise, we develop a multi-model ML approach with two main goals: (a) build a reliable and accurate relationship between the respondents' recent travel data on SM and their influence on future trip planning; (b) identify significant predictors that may affect future trip planning.

We collected primary data through a comprehensive survey of university students and young professionals in India, focusing on their SM usage, trip planning, content shared on SM, and SR on their recent trips. Using advanced ML models—logistic regression (LR), random forest (RF), and artificial neural networks (ANN)—alongside a Monte Carlo cross-validation technique and a data balancing method via oversampling, we developed reliable predictive models. The significant predictors are identified by using the average ranking of the variable importance scores obtained from the ML models. These averages are with respect to 100 Monte-Carlo replications and hence generate a robust ranking. This is a novel approach for identifying significant predictors, and hence a crucial methodological contribution. By empirically examining ASR as a factor in future trip planning, the current study offers a novel perspective through a rigorous analysis and actionable insights for both scholars and practitioners.

The remainder of the paper is structured as follows. Section 2 presents a brief literature review. Research hypothesis development and conceptual grouping of the features are discussed in Section 3. The proposed methodologies for data collection and multi-model ML development are presented in Section 4. Section 5 outlines an insightful summary of the data, ML model results, and how these results lead to different hypotheses. Finally, Section 6 concludes the paper with important remarks, theoretical and managerial implications, limitations, and a discussion on the future scope of work.

**2. Literature review**

This section presents brief literature reviews on the relevant topics, i.e., user generated SM content (UGC), the influence of SM on tourism, SR on trips, current trends in SM usage for travel planning, and the usage of ML models in tourism.

*User Generated Content (UGC):* Daugherty et al. (2008) define UGC as media created by the public rather than paid professionals and mainly shared online. UGC on SM significantly influences the travel experience (Dedeoglu et al., 2020). UGC influences travel across three stages: pre-travel, during, and post-travel (Tham et al., 2020; Mahaptra et al., 2025). Tourists seek UGC to plan specific and broad vacation activities (Yoo & Gretzel, 2011). Research has explored UGC on platforms like Facebook (Li et al., 2023), Twitter (Huang et al., 2020), YouTube (Chang, 2022), Instagram (i Agusti, 2021; Volo & Irimiás, 2021), and Snapchat (Jeffrey et al., 2022). Roma and Aloini (2019) compared UGC across Facebook, Twitter, and YouTube. In general, UGC is determined by the content that travelers choose to produce and post on their SM networks. Since UGC depends on what tourists share, analyzing SM content is essential to understanding travel-related content sharing and its link to behavior across all travel stages. Further research can help unpack the nature of self-generated SM content to investigate post-travel behavior to gauge pre-travel planning in the future. The current study aims to unpack this building on the existing literature of UGC.

*Tourism & SM Influence:* Since tourism travel is based on consumer behavior and decision-making, social influence—especially through SM—is inevitable. Current and prospective tourists engage with tourism both digitally and physically, with SM reviews shaping behaviors such as trip planning, on-site consumption, and post-trip evaluation (Sotiriadis, 2017). Hu and Olivieri (2020) describe tourists' SM accounts as customer-owned touchpoints that, while uncontrolled by organizations, guide other consumers' journeys. Extensive literature highlights the crucial role of SM in socially influenced tourism (Mittal et al., 2022; Boley et al., 2018; Munar & Jacobsen, 2014). Sharing tourism experiences on SM is also influenced by how the user positions their belongingness to a certain social group, ranging from following a common trend or standing out in terms of identification (Kang & Schuett, 2013; Eckhardt et al., 2015; Dinhopl & Gretzel, 2016; Canavan, 2017) to garnering social capital through their social network (Chen et al., 2021). Assuming SM users are tourists, their shared content—including photographs and received responses—offers insights into SM's influence on travel behavior (Antunes et al., 2018). Prospective research, like the current study, can investigate the SM influence in terms of self-generated content, such as photographs and posts, as well as other user generated content (UGC), such as received responses, to examine the nature of social influence in travelers' tourism intention to extend existing research in tourism literature.

*Social Return (SR):* Conceptually, SR provides an evaluation of the travelers' shared experiences along with a larger sense of validation by their social networks. This expected evaluation emerges in the form of likes, positive remarks, shares, etc. (Boley et al., 2018; Moran et al., 2018; Boley et al., 2023). Since SM is a widely used social platform for tourists to share their travel experiences (Mittal et al., 2022), tourists anticipate feedback on their self-generated content on SM platforms. Moreover, based on the responses from others (UGC), the experiential value of the tourist is further enhanced (Bigne et al., 2020). In addition to enhanced experiential value, tourists also form a sense of attachment to memorable experiences (Vada et al., 2019; Sakshi et al., 2020) at specific destinations, making their personal lives more memorable. The positive, memorable experiences also serve as powerful word-of-mouth (WOM) for themselves and other potential tourists (Mittal et al., 2022). Therefore, SR can impact the future travel planning of both the self and the other, making the previous travel more memorable (Wu et al., 2023; Bhogal et al., 2024). Drawing from the concept of Social Return (SR) as discussed in the literature (Boley et al., 2018; et al., 2023), SR appears to be linked to various behavioral factors related to travel planning, including positive or negative attitudes, social norms, and perceived control over behavior among prospective travelers. To assess the SR on intention to visit, our study has measured it through the reported number of accumulated likes as well as the types of reactions/responses received on the SM content shared by the traveler from recent short-term domestic trips. Therefore, the current study builds on existing literature of SR and enhances it further by conceptualizing and measuring a nuanced yet critical aspect of leisure travel.

*Current trends in SM usage on travel planning:* Although short-form SM influencers, videos, reels, and AI-based recommendations are outside the scope of this study, they represent notable trends in travel planning. Mahapatra et al. (2025) highlight the surge in SM influencers influencing tourists across all travel stages—pre-travel, during, and post-travel (Tham et al., 2020). Platforms like TikTok, YouTube, and Bilibili (Du, 2020; Liu et al., 2023; Xu et al., 2020) have popularized short videos, shifting travel information from traditional to SM channels and from textual to visual or multimodal content. Zheng et al. (2022) note that short videos create immersive experiences that foster tourism intentions. Additionally, travelers increasingly adopt AI-based recommendation systems to simplify information search and receive personalized trip plan (Shi et al., 2020).

*ML techniques in tourism studies:* Various ML models have been used in the current tourism literature. The applications of these models are diverse and effective. For instance, Yin and Jung (2024) use ML models to analyze the causes of tourists' emotional experience related to tourist attractions, and Zhang and Tang (2022) use RF models to identify attractive tourism spots. Moreover, Núñez et al. (2021) use a logistic regression model to find the determinants of tourism expenditure in Mexican households. To enhance tourism-related economic efficiency, Tian and Tang (2025) use ANN models. In addition to the studies mentioned above, the study by Núñez et al. (2024) provides a systematic review of ML models in tourism literature. A multi-model methodology, as executed in the current study, may enhance an empirical investigation that deals with young travlers' trip-planning behavior w.r.t. shared SM content from past trips.

## 3. Hypotheses development

To investigate our research question, we formulate five hypotheses to test the relationship between the influence of SR accompanied by four social constructs that are not directly observable but can be quantified using various related questions. We now briefly discuss the theoretical support for the hypotheses tested, along with the conceptual bucketing of the predictors.

Travelers' trip-planning behavior is shaped by social capital—the ability to access resources within social networks (Lam et al., 2024). Access to travel information and perceived social norms are influenced by SM profiles, which are characterized by privacy settings, connection count, activity level, access modalities, and usage frequency. Privacy reflects awareness of one's reference group—typically family and friends—which in turn shapes sharing behaviors. As Oliveira et al. (2020) note, potential travelers undergo an identification process to align with their social group, shaped by social influence (Kang & Schuett, 2013) and social capital (Chen et al., 2021). Understanding SM activity becomes crucial given the focus of this study on sharing content on SM that is both self-generated and created by others. India leads with 383.5 million Facebook users (Statista, 2025a), while Facebook and Instagram have 3 billion and 2 billion monthly active users globally (Statista, 2025b). These features facilitate the investigation of the relationship between the components of SM profile constituents and SR and trip-planning behavior in the following hypothesis:

*H1: Travelers' future trip-planning behavior depends on their SM profile constituents.*

Travelers' sharing behavior is reflected in SM content, including posts, discussions, and engagement frequency. Oliveira et al. (2020) identify two dimensions of this behavior: pre- and post-travel. Pre-trip sharing not only seeks information but also fosters interaction, reinforcing identification within social groups (Kang & Schuett, 2013). Discussions about the trip during the planning stage act as digital word-of-mouth, influencing travelers' choices of destinations (Boley et al., 2018; Boto-García et al., 2022). Friends, family, and broader networks are often used as reference groups in these discussions, which enhances intention and decision-making (Boto-García et al., 2022). SM sharing behavior includes various dimensions: attitude (willingness to share), decision (whether to post during travel), type (what is shared), and privacy (audience control). Together, these form the construction of SM content-sharing behavior, which this study investigates in relation to future trip planning through the following hypothesis.

*H2: Travelers' SM content-sharing behavior influences their future trip-planning behavior.*

Travelers are identified by their trip details and characteristics, including the number of recent trips, the length of the trips, the type of trips, and the rationale behind the selection of a certain type (Pratt et al., 2023). Active travelers are understood in terms of their travel style or the nature of their "travel party" -- independent or package travel (Hyde & Lawson, 2003). In our context, the more important factor is whether or not the participants have traveled in the past few years, regardless of their travel style. Further, current research (Su et al., 2024) suggests that trip details such as frequency, duration, trip type, etc., can influence tourists' trip satisfaction and revisit intentions. This implies that information about a traveler's most recent trip can act as a precursor to future trip-planning. For the purpose of this study, we have thus defined the concept of active travelers by assessing the length of travel, the number of trips, and the type of travel (leisure or not). We propose the following hypothesis on the active traveler's trip details.

*H3: Travelers' recent trip details influence their future trip planning.*

Tourists' self-consciousness is closely tied to how they share travel experiences through photography, their awareness of impression management (Rosenberg & Egbert, 2011; Li & Wan, 2025), and the tourist gaze (Belk et al., 2011; Urry & Larsen, 2011; Li & Xie, 2020). This self-awareness shapes decisions around trip photography. Sharing travel photos on social media (SM) also provides gratification (Liao et al., 2021), serving as both self-presentation and a means to attract social capital that may influence future UGC sharing (Munar & Jacobsen, 2014). In our study, trip photography encompasses camera usage as well as the quantity, kind, and sharing of photos. Tourist photography follows collective norms and imaginaries (Urry & Larsen, 2011) and is viewed as a tourism performance (Giovanardi et al., 2014; Gholamhosseinzadeh et al, 2021), where the experience is shaped by conventions like taking pictures. Through photo-sharing, travelers seek authenticity and escape from routine (Han & Bae, 2022), engaging in both object-based and experiential authenticity (Kontogeorgopoulos, 2017). These practices can influence future trip planning by reinforcing motivations for authentic travel experiences. Therefore, we propose the following hypothesis to test the relationship between trip photography and future trip planning behavior.

*H4: Travelers' trip photography demeanor influences their future trip planning.*

According to Boley et al. (2018, p. 120), "social return can be conceptualized as the amount of positive social feedback that one's social media posts will generate." Positive social feedback is thus divided into three categories: the number of likes received on the photographs, the responses received, and the expected reactions to the posts. Boley et al. (2023) also argue that the SR can serve as a significant predictor of a tourist's intention to visit a destination and other factors in the travel experience. According to Zheng et al. (2024), travelers' self-generated content on travel-related experiences generates positive emotion both for the observer and the publisher. This shared content then drives user engagement, both in the form of personal and interactive engagement behaviors. Therefore, rather than encouraging more user engagement, the shared social media content serves a motivation for future travel planning, positive emotions, and positive eWOM, even in the manner of affective reactions (Maiberger et al., 2024). Through this construct of feedback on the posts, we aim to gauge the awareness of SR while planning for future trips. In our study, the overall feedback is measured through the reported number of expected responses (likes, comments, reactions, etc.), the number of accumulated likes, and the types of reactions received on the shared content from recent trips. Hence, we propose the following hypothesis:

*H5: Travelers' overall feedback garnered on recent trips influences future trip planning.*

To quantify these five constructs (i.e., SM profile constituents, SM content-sharing behavior, trip details, trip photography, and feedback on SM posts), we collected data on 24 items/predictors along with demographic details such as age and gender (see Figure 1).

## 4. Proposed methodology

This section starts with a detailed discussion of the data collection and data cleaning process adopted for this study. Next, we present popular machine learning models with two objectives: (a) build a reliable and accurate relationship between the respondents' recent travel data on SM and their influence on future trip planning; (b) identify significant factors that may affect future trip planning. Various oversampling methods are investigated to mitigate the imbalance in the data, and a popular cross-validation technique is used to ensure the reliability of the predictive model.

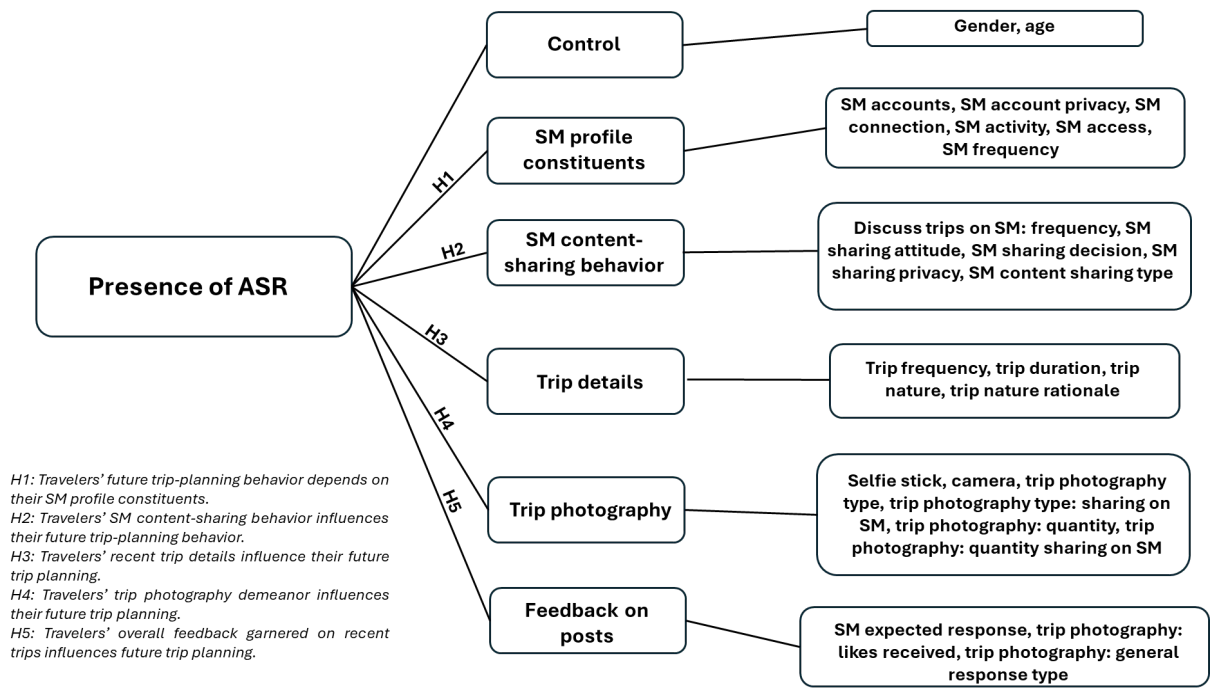

*Fig 1. Conceptual relationship between the presence of ASR, constructs, and predictors.*

## 4.1 Data collection

We designed a comprehensive two-stage survey for this study. Participation in the data collection was completely voluntary, and consent was obtained. Our target was to reach out to the age group of 18-50 who are active on SM, can independently decide when and where to travel, and have adequate financial support to do so. We floated the English-language survey to university students, professionals, executives, and acquaintances in our professional and personal networks in India, who represent the current urban population of India that can afford leisure travels without any financial constraint. People in these networks forwarded the survey to their friends and networks consisting of a similar population in terms of financial independence and decision-making attributes.

The survey questionnaire was designed in Qualtrics and circulated via email and WhatsApp to the target audience. First, a pilot survey was released on September 7, 2023, and was open till September 13, 2023. Out of 194 submissions received, 161 responses were considered for preliminary analysis since only those respondents had either Facebook (FB) or Instagram accounts. Based on the feedback, we made a few modifications, like minor linguistic updates, changed options for some questions, and removed a few redundant questions. We released the final survey with 30 questions on September 20, 2023, and collected responses till October 20, 2023. This time, we received a total of 783 submissions, of which 656 were valid. Since many respondents had left a few segments blank, and 61 participants had no SM accounts, the number of valid analyzable samples decreased to 318.

The information on the highest level of education, current occupation, and the state of residence (Q3-Q5) was used to validate the representativeness of the respondents in the target population and whether they could afford such trips. The remaining questions were divided into the target variable (Q30) and 26 predictors (Q1-Q2, Q6-Q29) (see Supplementary Material A for the full questionnaire with basic summary statistics).

(a) Data on gender and age (Q1-Q2) of the respondents were considered as control.

(b) SM engagement was captured through questions (Q6-Q19) on how active they are on SM platforms, what is the number of friends/connections, what they post, how much they share when they post content on SM, whether their SM accounts are public or private, how often they visit the platform, etc.

(c) Q20 records the number of trips between January 2022 and August 2023. Q22 indicates that the trips were mostly leisure (94.96%). We asked trip-wise details (up to a maximum of 4 recent trips as per their recall) on the duration of the trips, the nature of the trips, what they posted from their trips on social media, what kinds of pictures they took and shared, the number of likes and comments received on their self-generated content on SM, etc. Furthermore, data from different trips (between one and four as provided by the respondents) were combined for each question (Q21-Q30) in this section as per the max-pooling or sum-pooling approach (see Supplementary Material A for details).

An insightful summary and plots for a few questions are presented in Section 5.1.

### 4.2 Predictive modeling

In this section, we present three state-of-the-art ML models for accurately predicting the target variable, $Y_i$ - "Presence of ASR" reflecting influence on future trip planning behavior for the i-th respondent, using $X_i$, which consists of gender and age group (as control), variety of SM activities, trip details, and SR generated from the SM posts.

The class of admissible predictive models depends on the data type of the target variable. Since $Y_i$ is dichotomous (i.e., coded as 0 if there would be no change in the future trip planning and 1 otherwise), either a binary classifier or a prediction model that allows binary response variables needs to be used. We have selected three popular predictive models from a long list of statistical and machine learning models that can be considered for this exercise (see Hastie et al. 2009). Further discussion on specific reasons for choosing these models is presented in the next section and summarized in Table 1.

### 4.2.1 Machine learning models

*Logistic regression (LR) model* is the most popular and conventional parametric model in the class of generalized linear regression models for binary responses. One of the reasons for its popularity is the interpretability of regression coefficients and the ability to assess their impact on /significance of the response, the second research objective of this study (i.e., the identification of significant factors that may affect future trip planning).

In the LR model, the log of odds of $P(Y_i=1)$, i.e., the probability of changing the trip plan due to the presence of ASR, is considered as a linear model,

$$\log\left(\frac{P(Y_i=1)}{1-P(Y_i=1)}\right) = \beta_0 + x_i^T \beta,$$

where $x_i$ and $\beta$ are 26-dimensional vectors with respect to the output of the predictors, and the corresponding regression coefficient, respectively (see Kutner et al., 2005, and Hastie et al., 2009 for detailed methodology). For model implementation, we used "glm()" function in R software with family="binomial" and default values for other arguments like the number of iterations, starting values for likelihood optimization, etc.

*Random forest (RF) model*: Decision trees are a well-established non-parametric alternative for regression and classification predictive models in the ML literature and span a wide spectrum of applications, including tourism (e.g., Zhang & Tang, 2022). Leo Breiman pioneered the idea of tree-based models in 1984, which later evolved into "Random Forest" (see Breiman (2001) for detailed methodology). An RF model is an ensemble of decision trees constructed over bootstrap samples of the data, where every node is split further using a small random subsample of the predictors. The idea of averaging over several trees makes the RF model very robust and reliable.

For building an RF model, we used the R function "randomForest()" with hyperparameters ntree=1000, which represents an ensemble of 1000 random binary decision trees, and mtry=5 (approximately equal to the square root of the number of predictors – default value of the software), which corresponds to using five random predictors to find the best split at each node. The rest of the arguments, like the minimum node size for splitting, depth of the tree, misclassification error, complexity parameter value, etc., were set at the default values.

*Artificial neural network (ANN) models* represent a class of flexible non-linear predictive models that have gained immense popularity in various domains like drug discovery, financial trading, cybersecurity, manufacturing, IOT, tourism, simulator building, etc. ANN-based models serve as the key element in many AI models as well.

The basic idea behind constructing an ANN model is to create a nested network of variables, from the set of observable inputs to the output via several layers of latent variables coupled with activation functions. This provides massive flexibility to the model that can capture complex relationships. Compared to several other ML models, ANN typically gives highly accurate predictions but requires a relatively larger dataset as the number of parameters multiply with the number of hidden layers. We use a simple ANN model with only one hidden layer to build the predictive model using the "nnet()" function in R. Most of the hyperparameters were set at the default values, e.g., the maximum number of iterations for the estimation of regression coefficients was set to 100, and the sigmoid activation function was used. However, the number of latent variables in the hidden layer was treated as a tuning parameter, and Section 5.2 elaborates on its estimation.

The statistics / ML modeling literature contains a plethora of predictive models, and for a given problem, one cannot use all possible models, but there is no unique choice either.

We used this multi-model ML approach to cover the spectrum of admissible predictive models for our study. It spans a classical approach (LR) for benchmarking, to the most reliable predictive model (RF), and a highly accurate prediction methodology (ANN). Table 1 summarizes the key features of these three predictive models.

*Table 1. Comparison of the salient features of the three predictive models used in this paper. Alt text: The table presents key features of the three predictive models used here.*

|  | Important features | Purpose / Reason |
| --- | --- | --- |
| Logistic regression | Simple parametric model; closed-form predictive model; facilitates significance assessment of the predictors | For benchmark comparison |
| Random Forest | Nonparametric model with ensemble approach | Gives a reliable prediction |
| Artificial Neural Network | Flexible nonlinear model; basis for various AI models | Gives high prediction accuracy |

**4.2.2 Data imbalance**

Classification models with binary and/or categorical responses suffer from data imbalance, particularly in the target variable. In this study, we have 91 observations with Y=1 and 227 data points with Y=0. Such an imbalance in the data often leads to poor model fits and, hence, inaccurate predictions. Chawla et al. (2002) proposed a popular technique called SMOTE (*Synthetic Minority Over-sampling Technique*) to address this issue by generating synthetic samples using k-nearest neighbours of the minority class, and then fitting the model on the pooled data. Several extensions and modifications were subsequently developed to address and improve the drawbacks of SMOTE. Refer to the R library "imbalance" (Cordón et al., 2018) for the implementation of various algorithms, including ADASYN (adaptive synthetic sampling approach), MWMOTE (majority weighted minority oversampling technique), RWO (random walk over-sampling), RACOG (probabilistic oversampling approach) PDFOS (pdf estimation-based over-sampling), and NETER (filtering of over-sampled data using non-cooperative game theory). In this paper, we have compared the usage of various methods for addressing the data imbalance issue.

**4.2.3 Prediction reliability**

Enhancing the prediction accuracy by introducing latent variables in ANN, or augmenting synthetic observations to address data imbalance, often leads to model overfitting. A standard practice in the modeling literature is to split the data into train and test sets. The ML models are fitted on the training data and then validated on the test data. A model is said to be an overfit if the training error is much smaller than the test error, whereas a model is called reliable if the prediction accuracies in the training and test sets are comparable. Although the idea of train-test split provides an efficient tool for the reliability assessment of the model, the randomness in data splitting may lead to biases, which in turn can result in differences in the prediction accuracy values and the set of significant/important factors identified by the ML models (the two main research objectives of this paper). Cross-validation is commonly used to overcome this train-test split bias. The key idea is to generate multiple copies (say, M) of train-test splits, fit models on the train sets, obtain predictions on both train and test data for each of the M sets, and then compare the average train and average test accuracies. A robust model should yield a ratio of average train and average test accuracies close to one. Popular cross-validation methods are k-fold, Monte Carlo, stratified, leave-one-out, bootstrapping, etc. (Hastie et al., 2009). There is no universally accepted ranking among different cross-validation methods for all applications. The results presented in Section 5 are based on Monte Carlo cross-validation approach.

## 5. Results

This section covers three topics: (a) insightful summary statistics of the survey data, (b) the results from the ML model fits, and (c) a discussion of the results concerning the overall research objective and hypotheses outlined in Section 3.

### 5.1 Descriptive statistics

Although we followed a convenience sampling approach, the respondents hail from a wide spectrum of adults in India. Figure 2 summarizes the data distribution as per the four variables: gender, age group, education level, and occupation. In the data considered for modeling, 127 respondents are female, and 191 are male. The age group 25-35 has the largest representation of 132 (out of 318) samples. Many of the respondents have a bachelor's or a master's degree, and most respondents are either students or service personnel in the industry.

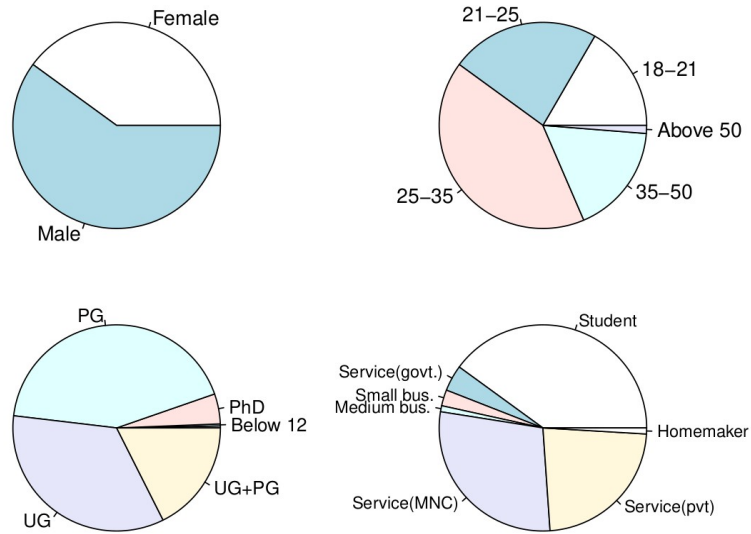

*Fig 2. Distribution of 318 respondents from India for gender, age, education level, and occupation.*

A closer look at the distribution of age and occupation (Table 2) suggests that our respondents were financially capable of supporting leisure travel. The largest occupational group comprised service-sector employees in MNCs, followed by those in private services and students, underscoring their financial independence. Moreover, as understood from both Figure 2 and Table 2, most respondents fell within the 18–50 age range, representing both Gen Y and Gen Z—the target demographic of this study—who are characterized as young urban travelers frequently undertaking short leisure trips.

*Table 2. Distribution of the respondents' counts with respect to age group and occupation.*

| Occupation | Age group | | | | |
|---|---|---|---|---|---|
| | 18-21 | 21-25 | 25-35 | 35-50 | 50 and older |
| Homemaker | 0 | 1 | 2 | 0 | 0 |
| Medium to large-scale business | 0 | 1 | 2 | 0 | 0 |
| Service (govt.) | 0 | 3 | 6 | 4 | 0 |
| Service (MNC) | 0 | 14 | 48 | 28 | 1 |
| Service (private) | 1 | 9 | 40 | 20 | 3 |
| Small-scale business | 0 | 3 | 3 | 2 | 0 |
| Student | 52 | 43 | 31 | 1 | 0 |

We also looked at the geographical coverage of respondents in India and found that Uttar Pradesh, Delhi (north), Madhya Pradesh (center), Karnataka (south), Maharashtra (west), and West Bengal (east) have substantial participation in the samples. Although not shown in Table 2, except for one, all participants were college-educated: 110 held undergraduate degrees, 56 held integrated UG–PG degrees or diplomas, 136 held postgraduate degrees, and 15 held PhDs. This highly educated profile, which has also been broadly indicated in Figure 2, suggests that young Indian travelers in our sample are likely to engage in informed decision-making regarding social media use and travel planning. In summary, the collected data is a representative sample that can be utilized to research the target population of urban SM users who participate in leisure trips at their own discretion. Supplementary Material A presents the state-wise distribution of respondents' representation.

A high number, 216 respondents, had both FB and Instagram accounts, 82 reported having only Instagram, and 20 had only FB. This trend is in line with what the current reports (NapoleonCat, 2025a, 2025b) about these two platforms in India. Almost 65% of respondents reported their SM accounts to be private with regard to content sharing. The distributions of frequency of SM apps/website visits (left panel) and number of SM connections (right panel) are shown in Figure 3.

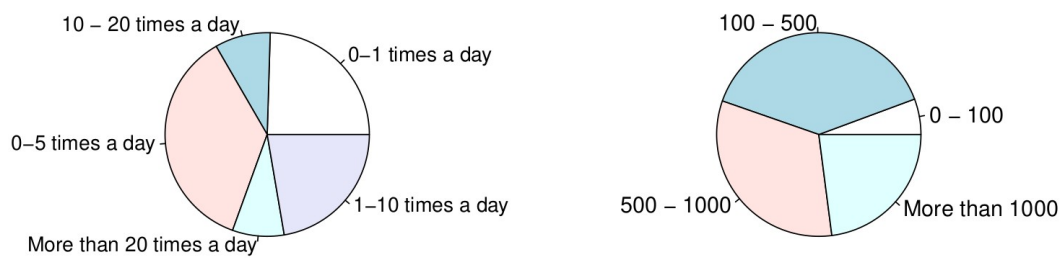

*Fig 3. Distribution of respondents with respect to the number of SM connections (right panel) and the frequency of SM account visits (left panel).*

The distribution of the number of trips (depicted in Figure 4) shows that a few respondents made up to 25 trips in this study period (January 2022 – August 2023), whereas most had made 2-4 trips. Our data (Q21 and Q22 in Supplementary Material A) shows that most of these trips were leisure (94.96%) and short-term tourism travels of 2-4 days or 7-10 days. This indicates that the respondents had adequate time to avail such trips.

When asked, "How often do you discuss your trips with connections on social media accounts?" 29% of the respondents said never, almost 53% discuss trips sometimes, and 18% often or very often. These provide compelling reasons for us to investigate it further and statistically validate our research questions using the proposed multi-model approach. That is, whether SM engagement, willingness to share content on SM, and their feedback on previous travel experiences can influence the participant's decision to change future trip planning.

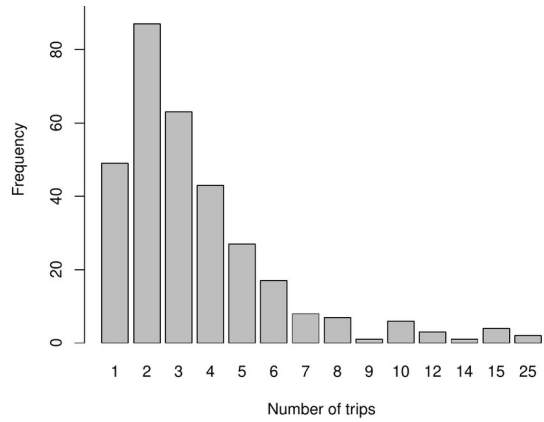

*Fig 4. Distribution of the number of trips by the respondents.*

## 5.2 ML model results

This section presents the model fit results of different predictive models used in building the proposed multi-model framework. First, the LR model fit-based results are used to select an appropriate oversampling method to balance our data, and then we use this balanced data to compare the average predictive power of the three ML models via Monte Carlo cross-validation. Since these predictive models are binary classifiers, the prediction performance is assessed through various goodness of fit (GOF) measures obtained from the confusion matrix. A good model should maximize the following measures:

- Overall accuracy (OA): proportion of correctly predicted responses,
- Sensitivity: proportion of correctly predicted 1s out of all true 1s,
- Precision: proportion of correctly predicted 1s out of all predicted 1s,
- Specificity: proportion of correctly predicted 0s out of all true 0s,
- F1-score: harmonic mean of precision and sensitivity,
- AUC: area under the receiver operating characteristic (ROC) curve.

The LR model fitted to the "original" data (91 observations with Y=1 and 227 data points with Y=0, i.e., a total of 318 observations) shows poor results (see Table 3). Particularly, the OA value is reasonably high, specificity is much higher, but the sensitivity and F1 scores are very low (see the first column of Table 3). Such a phenomenon is typically a consequence of the imbalanced data. We used various oversampling techniques from the R library "imbalance" to generate synthetic samples to balance the two classes of response. The GOF measures of the LR model fit on the balanced data obtained via different oversampling methods are summarized in Table 3. It is clear from Table 3 that the SMOTE method is the poorest of the lot, and MWMOTE, RWO, and PDFOS are relatively similar, but PDFOS is marginally better than the others. Therefore, we use PDFOS (PDF estimation-based oversampling) – based balanced data (with 227*2=454 observations) for the subsequent ML model fitting exercise.

*Table 3. Goodness of fit measures for LR model fitted to the original (imbalanced) and balanced data through various oversampling methods.*

|  | Original | SMOTE | MWMOTE | RACOG | RWO | PDFOS | NEATER |
|---|---|---|---|---|---|---|---|
| OA | 0.75 | 0.71 | 0.72 | 0.66 | 0.75 | 0.76 | 0.75 |
| Sensitivity | 0.30 | 0.47 | 0.70 | 0.65 | 0.70 | 0.72 | 0.60 |
| Precision | 0.64 | 0.65 | 0.73 | 0.66 | 0.78 | 0.78 | 0.74 |
| Specificity | 0.93 | 0.85 | 0.74 | 0.66 | 0.80 | 0.80 | 0.85 |
| F1 score | 0.41 | 0.55 | 0.72 | 0.65 | 0.74 | 0.75 | 0.66 |
| AUC | 0.74 | 0.75 | 0.81 | 0.72 | 0.84 | 0.84 | 0.80 |

We now proceed with the cross-validation aspect of the model-building step to obtain reliable predictions. Suppose the full data with N observations is split into two parts: a train set with N1 data points and a test set with N2 observations. Typically, the train-test proportion varies between 70 - 30 and 80 -20. In this paper, we have used a 75 – 25 split for generating the train and test sets. Since the inclusion of hidden layers with latent variables in neural network-based models and oversampling the minority class may lead to overfitting the training data, we apply Monte-Carlo cross-validation to avoid overfitting.

We fit the LR model to the training data and compute different prediction accuracies based on the confusion matrix. While fitting the model, we applied the forward selection method with the AIC (Akaike's information criterion) to keep the relevant predictors and eliminate redundant ones. This will be helpful in addressing our second research objective to identify important predictors that may affect future trip planning. We also computed the corresponding prediction accuracies for the test data using the fit obtained from the training data. This process was repeated 100 times with random train-test splits. Table 4 presents the average prediction accuracies for both the original (imbalance) data and after balancing the data via PDFOS-based oversmoothing.

Table 4 shows that the average prediction accuracies for the balanced data are consistently high and stable as compared to the imbalanced (original) data. Furthermore, the average train accuracies are much closer to test accuracies (ratios close to 1) for the balanced data as compared to the unbalanced data. In summary, the LR model is giving 72% overall accuracy, 70% F1-score, AUC = 0.79, and sensitivity of 67%. Significance of predictors under this model will be discussed along with other models in Section 5.3.

*Table 4. Train and test prediction accuracies (averaged over 100 Monte Carlo replications) for the LR model fitted to both the original (imbalanced) and balanced data obtained via the PDFOS method.*

|  | Imbalanced data | | | Balanced data | | |
| --- | --- | --- | --- | --- | --- | --- |
|  | Train | Test | Ratio | Train | Test | Ratio |
| OA | 0.74 | 0.68 | 1.08 | 0.76 | 0.72 | 1.05 |
| Sensitivity | 0.24 | 0.14 | 1.71 | 0.71 | 0.67 | 1.06 |
| Precision | 0.60 | 0.37 | 1.62 | 0.79 | 0.74 | 1.06 |
| Specificity | 0.94 | 0.90 | 1.04 | 0.81 | 0.76 | 1.06 |
| F1 score | 0.34 | 0.20 | 1.7 | 0.75 | 0.70 | 1.07 |
| AUC | 0.72 | 0.58 | 1.24 | 0.85 | 0.79 | 1.07 |

For the RF model as well, we followed a similar approach, i.e., compared the prediction accuracies between the original (imbalanced) and balanced data obtained using the PDFOS-based oversampling method. Data was split into train and test as per a 75-25 ratio, the RF model was fitted on the train data, and then predictions of both train and test sets were used for computing the GOFs. Each RF model represents an ensemble of 1000 random classification trees, with the minimum node size kept at 10 as the pruning parameter. Table 5 presents the average GOFs from 100 replications. Clearly, the balancing step has significantly stabilized the GOF measures. Since the RF model is known for its reliable predictions, the ratio of average train and average test accuracies for the balanced data is very close to one, and overfitting is not observed here – a compelling reason for including this model in building our robust predictive framework.

*Table 5. Train and test prediction accuracies (averaged over 100 Monte Carlo replications) for the RF model fitted to both the original (imbalanced) and balanced data obtained via the PDFOS method.*

|  | Imbalanced data | | | Balanced data | | |
| --- | --- | --- | --- | --- | --- | --- |
|  | Train | Test | Ratio | Train | Test | Ratio |
| OA | 0.71 | 0.71 | 1.00 | 0.77 | 0.76 | 1.01 |
| Sensitivity | 0.09 | 0.10 | 0.90 | 0.69 | 0.69 | 1.00 |
| Precision | 0.46 | 0.55 | 0.84 | 0.82 | 0.81 | 1.01 |
| Specificity | 0.95 | 0.96 | 0.99 | 0.85 | 0.84 | 1.01 |
| F1 score | 0.17 | 0.17 | 1.00 | 0.75 | 0.74 | 1.01 |
| AUC | 0.98 | 0.61 | 1.60 | 0.99 | 0.84 | 1.17 |

A quick comparison between Table 4 and Table 5 shows that the RF model (test accuracies range in 69%-84%) is more accurate than the LR model (test accuracies range in 67%-79%). This is expected as the RF model is non-parametric and a significantly more powerful predictive tool than a generalized linear model (GLM). We now explore the prediction performance of the ANN model, which is known for its complexity but also has the ability to give very high prediction accuracy. ANN models may contain hundreds to thousands of parameters (depending upon the number of latent variables) and require big data for model fitting. Since we have only 454 observations, we investigate the usage of an ANN model with only one hidden layer with "k" latent variables. We conduct a small simulation study to find a suitable choice for "k". The overall accuracy (OA) values of the model fitted to the balanced data (using PDFOS method) were compared for different values of "k" in the range of 1 to 15. Figure 5 presents the distribution of OA values with respect to "k" for 50 Monte Carlo replications.

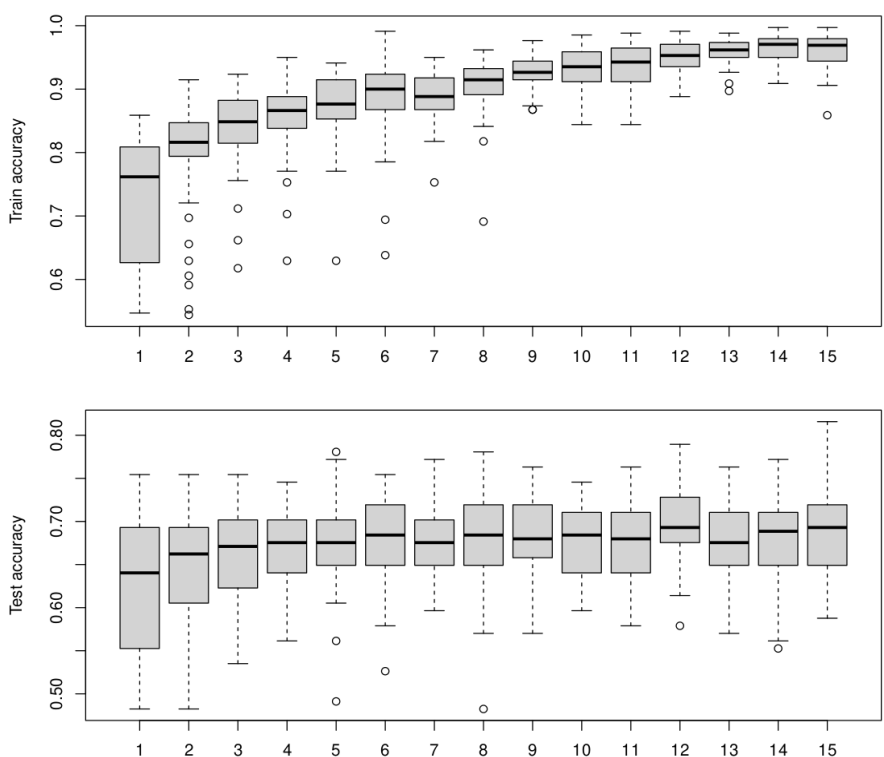

*Fig. 5. Distribution of train- and test-OA for the ANN model fitted to the balanced data with one hidden layer and different numbers of latent variables.*

The following two findings emerge from Figure 5: (a) the Train-OAs are significantly higher than the Test-OAs, and (b) the Train-OAs increase with the number of latent variables, whereas the Test-OAs are somewhat stable. For the subsequent ANN results in this paper, we used one hidden layer with ten latent variables, as they gave more than 90% average training accuracy in approximately half of the replicates. We now follow the same Monte Carlo cross-validation approach as in LR and RF models and compare the prediction accuracies of ANN model fits for both balanced and imbalanced data. The results in Table 6 are averaged over 100 Monte Carlo replications.

*Table 6. Train and test prediction accuracies (averaged over 100 Monte Carlo replications) for the ANN model fitted to both the original (imbalanced) and balanced data obtained via the PDFOS method.*

|  | Imbalanced data | | | Balanced data | | |
| --- | --- | --- | --- | --- | --- | --- |
|  | Train | Test | Ratio | Train | Test | Ratio |
| OA | 0.92 | 0.63 | 1.46 | 0.93 | 0.67 | 1.38 |
| Sensitivity | 0.83 | 0.33 | 2.51 | 0.89 | 0.64 | 1.39 |
| Precision | 0.89 | 0.35 | 2.54 | 0.96 | 0.69 | 1.39 |
| Specificity | 0.95 | 0.75 | 1.27 | 0.96 | 0.71 | 1.35 |
| F1 score | 0.85 | 0.33 | 2.57 | 0.92 | 0.66 | 1.39 |
| AUC | 0.95 | 0.57 | 1.67 | 0.96 | 0.72 | 1.33 |

It is clear from Table 6 that balancing the data helps in stabilizing the difference between train and test accuracies, but the overfitting issue is very clear as the train/test ratios are significantly bigger than one. Moreover, a quick comparison with RF and LR models suggests that the test accuracies are relatively lower (in the range 64%-72%), i.e., less accurate predictions for the unused data.

## 5.3 Discussion

This section starts with a quick comparison of the three predictive models with respect to the prediction accuracies averaged over 100 Monte Carlo replications as reported in Tables 4 - 6. Subsequently, we delve into the discussion on the identification of important predictors that influence the respondents' trip planning behavior.

*Table 7. Comparison of prediction accuracy and reliability of the LR, RF, and ANN model-fits on balanced data.*

|  | Prediction Accuracy (test accuracy range) | Reliability (train/test ratio range) |
| --- | --- | --- |
| LR model | 67% - 79% | 1.05 – 1.07 |
| RF model | 69% - 84% | 1.00 – 1.01 (exception: 1.17 for AUC) |
| ANN model | 64% - 72% | 1.33 – 1.39 |

It is clear from Table 7 that the RF model gives the most accurate and reliable prediction for new data points. The LR model is not too different in terms of reliability, but with significantly less prediction accuracy. ANN model, on the other hand, is least reliable in terms of out-of-sample prediction. This addresses our first research objective – to build a reliable and accurate relationship between the respondents' recent travel data on SM and their influence on future trip planning

We now focus on the second research objective, i.e., the identification of important predictors via an innovative and robust approach of taking the average of the ranking criterion over 100 Monte Carlo replications. Table 8 compares the role of all 26 predictors in the three ML models (LR, RF, and ANN) fitted to the training set part of the balanced data. These results are based on 100 Monte Carlo replications with random train-test splits. The first column presents the variable names in the order they appeared in the questionnaire (see Supplementary Material A). The second column, "LR significance," presents the number of replications (out of 100) in which the variable was identified as significant as per the AIC criterion within the "forward selection framework". That is, the larger this number is, the more influence the variable has on the travelers' decision to change their trip-planning behavior. The last three columns contain the average importance ranking of the predictors for LR, RF, and ANN models, respectively. Here, the lower the ranks, the higher the importance of the variable is. Since these columns present the averaged ranks over 100 replications, decimal values and similar magnitudes are acceptable (e.g., 18.44 and 18.36 in the "LR importance" column).

*Table 8. Average importance ranking of all 26 predictors with respect to the three predictive models.*

| Variable names | LR significance | LR importance | RF importance | ANN importance |
|---|---:|---:|---:|---:|
| Gender | 24 | 14.74 | 23.51 | 13.89 |
| Age group | 20 | 15.99 | 15.86 | 12.49 |
| SM accounts | 21 | 18.36 | 20.06 | 12.55 |
| SM account privacy | 78 | 7.48 | 12.22 | 14.66 |
| SM connections | 25 | 18.44 | 16.17 | 12.68 |
| SM activity | 32 | 11.5 | 13.81 | 14.11 |
| SM access | 100 | 1.45 | 3.44 | 7.8 |
| SM frequency | 6 | 16.73 | 18.94 | 17.44 |
| Discuss trips on SM: frequency | 2 | 21.98 | 19.75 | 16.5 |
| SM sharing attitude | 99 | 3.03 | 10.49 | 9.66 |
| SM sharing decision | 95 | 5.81 | 18.32 | 9.47 |
| Selfie stick | 6 | 20.62 | 25.45 | 23.47 |
| Camera | 26 | 15.7 | 25.08 | 22.15 |
| SM sharing privacy | 71 | 13.7 | 11.79 | 16.54 |
| SM content sharing type | 2 | 22.11 | 22.04 | 12.8 |
| SM expected response | 4 | 22.38 | 1.75 | 12.89 |
| Trip frequency | 53 | 13.29 | 11.66 | 14.02 |
| Trip duration | 12 | 20.88 | 11.18 | 8.92 |
| Trip nature | 100 | 1.86 | 3.64 | 9.98 |
| Trip nature rationale | 42 | 13.39 | 2.58 | 16.44 |
| Trip photography type | 41 | 16.72 | 6.28 | 13.35 |
| Trip photo type sharing on SM | 64 | 13.02 | 9.41 | 11.54 |
| Trip photography: quantity | 43 | 12.65 | 9.61 | 14.24 |
| Trip photo: quantity sharing on SM | 21 | 12.07 | 7.92 | 9.09 |
| Trip photo: likes received | 96 | 4.49 | 6.46 | 5.7 |
| Trip photo: general response type | 67 | 12.61 | 23.58 | 18.62 |

A few notable findings from Table 8 are as follows. First, the rankings of the predictors as per the average variable importance are different for the three ML models. This is expected as the three predictive models are different by design, i.e., the LR model is a simple generalized linear regression model, the RF model is non-parametric and follows an ensemble technique, whereas the ANN model has several latent variables with non-linearity introduced via activation functions. Second, the top ten predictors identified by the three ML models as per the average importance ranking criterion cover all five hypotheses (Table 9).

*Table 9. Top ten predictors for the ML models identified by the average importance ranking criterion in Table 8.*

| LR model | RF model | ANN model |
|---|---|---|
| SM access (H1) | SM expected response (H5) | Trip photo: likes received (H5) |
| Trip nature (H3) | Trip nature rationale (H3) | SM access (H1) |
| SM sharing attitude (H2) | SM access (H1) | Trip duration (H3) |
| Trip photo: likes received (H5) | Trip nature (H3) | Trip photo: quantity sharing on SM (H4) |
| SM sharing decision (H2) | Trip photography type (H4) | SM sharing decision (H2) |
| SM account privacy (H1) | Trip photo: likes received (H5) | SM sharing attitude (H2) |
| SM activity (H1) | Trip photo: quantity sharing on SM (H4) | Trip nature (H3) |
| Trip photo: quantity sharing on SM (H4) | Trip photo type sharing on SM (H4) | Trip photo type sharing on SM (H4) |
| Trip photo: general response type (H5) | Trip photo: quantity (H4) | Age group |
| Trip photo: quantity (H4) | SM sharing attitude (H2) | SM account (H1) |

Third, a good predictive model should give somewhat uniformly distributed values of the average variable importance. This would imply consistency in the predictors' importance ranking across different Monte Carlo replications. Figure 6 presents the distribution of the last three columns of Table 8, i.e., the average importance ranking (over 100 Monte Carlo replications) for the LR, RF, and ANN models. It is clear from Figure 6 that the RF model gives the best results (closest to uniform distribution), whereas the ANN model shows a sharper peak near the middle, which implies significant variation in the predictors' importance ranking for different Monte Carlo replications. The overall inference of the three ML models drawn here is consistent with the findings of Table 7.

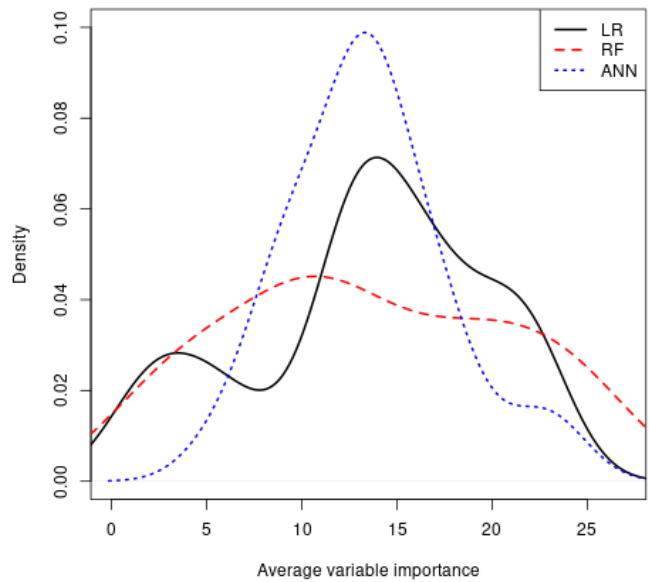

*Fig. 6. The distribution of average variable importance of all predictors for the three ML models.*

In summary, five predictors ("SM access," "SM sharing attitude," "Trip type," "Trip photography: quantity sharing on SM," and "Trip photography: likes received") are common in the list of top ten predictors identified by the three predictive models (as in Table 9). It is interesting to note that these five predictors support the five hypotheses listed in Section 3. That is, the dependence of SM profile constituents on future trip-planning behavior (H1) is supported by the significance of "SM access"; the influence of SM content sharing behavior (H2) is validated by the importance of "SM sharing attitude"; the influence of recent trip details (H3) is backed up by the identification of "Trip type" as an important predictor by most of the predictive models; the relevance of travelers' trip photography demeanor (H4) gets identified through the significance of "Trip photography: quantity sharing on SM"; and the influence of feedback garnered on the recent trips (H5) is captured via the identification of "Trip photography: likes received" as important by most of the ML models.

Table 10 summarizes the distribution of these five predictors with respect to different age groups in our dataset. We only present the first 4 age groups since the population of 50-above is negligible as compared to other categories. Please refer to Supplementary Material A for details on the coding and data distribution of values for each of these predictors.

*Table 10: Age-group-wise distribution of the five most important predictors identified by the three ML models.*

| Age group | SM Access [% of people with "App" on their phone] | SM sharing attitude [Median of general sharing on a scale of 1-5] | Trip type [% of leisure trip] | Trip photography: quantity sharing on SM [Median no. of photos on a scale of 1-16] | Trip photography: likes received [Median no. of likes across all trips] |
|---|---|---|---|---|---|
| 18-21 | 98% | 4 | 92% | 4 | 4 |
| 21-25 | 92% | 3 | 97% | 4 | 4 |
| 25-35 | 93% | 3 | 96% | 6 | 6 |
| 35-50 | 91% | 3 | 91% | 4 | 3 |

The predictors offer valuable insights with respect to different age groups. For example, the 18-21 population is most inclined to have SM applications on their phones, although all age groups are very inclined to have applications. Similarly, the same age group is most inclined to share content on SM, along with all the other age groups with high inclinations. Trip type is leisure for all the age groups, whereas the 35-50 population is relatively low for this kind of reported trip. The age group of 25-35 is most inclined to share trip photos and hence most inclined to receive likes on those as social media feedback. However, the 35-50 population remains low in terms of receiving likes on the posts.

## 6 Concluding Remarks

The overall goal of this paper was to investigate whether feedback on previous travel experiences related posts on SM, along with other aspects of a traveler's SM usage behavior, impact the participant's future trip planning. This was achieved by setting two specific research objectives: (a) build a reliable and accurate relationship between the respondents' recent travel data on SM and their influence on future trip planning; (b) identify significant predictors that may affect future trip planning. We also postulated five explicit hypotheses to validate our beliefs on the factors that might influence the future trip planning decision.

The findings in this paper are based on a representative sample of financially competent individuals (mostly students and young working executives) from urban areas of India who can afford leisure trips. The data collected for the investigation using an extensive online survey consisted of 318 complete and analyzable records. For the first objective, we addressed the imbalance in the data using state-of-the-art oversampling methods and developed an ML-based multi-model framework to build an accurate predictive model. The reliability of these predictive models was ensured via Monte Carlo cross-validation.

Furthermore, the second objective was achieved by identifying significant predictors from the LR model and computing variable importance scores from the RF and ANN models. Using the average ranking of variable importance (over 100 Monte-Carlo replications) for finding the significant predictors is an innovative methodological approach. Findings suggest that all the hypotheses are supported by each model (LR, RF, and ANN) through a subset of the predictors considered in the study, although not necessarily the same subset. The following features appear to have the most significant influence on the future trip planning decision, i.e., SM access on phone, consider SM when planning the trips, trip type, number of photos shared on SM, and the feedback garnered on the SM content posted from the recent trips. The RF model can accurately and reliably predict (with up to 84% average prediction accuracy) the participants' decision on future trip-planning behavior and hence suggest the significant presence of ASR.

*Theoretical implications*

In exploring the dynamics of travelers' SM behavior in trip planning, it is essential to consider the powerful influence of the novel social factor, ASR, on the decision-making processes. Some crucial concepts emerge in this investigation. First, so far, the notion of social return has either not been adequately addressed in the literature in relation to travel decision-making or has been touched upon to investigate how it might impact memorable tourist experiences and behavior intentions in terms of satisfaction (Mittal et al., 2022). Additionally, no significant study has been conducted where ASR is investigated empirically, utilizing Indian data and advanced machine learning techniques to establish a reliable, model-driven predictive approach. The methodology, along with the novel data set, shifts focus from memorable travel experiences and general consumption behavior of tourists to more SM-oriented behaviors. The significant predictors that emerge in the models are an important indication of how UGC should be called tourist-generated content (Antunes et al., 2018). Moreover, the customer-owned uncontrolled touchpoint (Hu & Olivieri, 2020) of tourists' social media accounts exhibits a tendency of strong ASR in trip-planning behavior as a part of the consumer journey. Therefore, the focus of this research extends beyond the empirical analysis of travel photographs, social media content, or general planning behavior. It seeks to examine the substantive presence of ASR within the discourse of trip-planning behavior, as methodologically shaped by the dynamic interrelationships among these constructs.

*Managerial implications*

The findings of this study also offer valuable insights, both in terms of actionable predictors and methodological inferences, for managers and marketers in the travel industry. The five key predictors of future trip planning to offer actionable insights for travel marketers are: "SM access on phone," "SM sharing attitude," "Trip type," "Number of photos shared on SM," and "likes received on the photos shared."

     First, the importance of "SM access on phone" highlights the need to focus on mobile-first platforms. Travel businesses should ensure their social media content is optimized for mobile use, with seamless booking and engagement features accessible through apps or mobile browsers. In this, businesses need to think more about features suitable for applications for the age group of 18-21 years and more about features suitable for browsers for other age groups. Second, "SM sharing attitude" points to the role of UGC in influencing planning behavior and highlights that travelers' willingness to share reflects not only their need for expression but also their engagement with the planning process. In order to promote the creation of narratives that can influence peers' plans, destination marketing brands should create experiences that are convenient for travelers to share and provide incentives for content creation. In this, managers and marketers should focus on the youngest age group, 18-21, to promote the sharing of destination-specific narrative content. Third, the role of "Trip type" suggests that recent travel experiences shape future plans. Marketers can use past trip data to tailor recommendations and design retention campaigns targeting specific traveler profiles. The age group of 35-50 may require more attention since their leisure trip participation is low compared to other age groups. Fourth, the significance of "Number of photos shared on SM" indicates that trip photography remains central to travel expression. By incorporating simple social-sharing technologies, destinations and service providers may encourage sharing and create photo-worthy surroundings. In this specific predictor, the age group of 25-35 needs more attention since they tend to share more photographs from trips. Finally, "likes received on photos shared" confirms that social feedback matters. Recognizing and amplifying popular user content can increase visibility and drive engagement from others planning trips. Additionally, the age group of 25-35 can be encouraged further since they show more inclination toward both sharing travel photos and eventually receiving more social media feedback. Travel marketers could leverage algorithms to highlight popular content created by fellow travelers.

For tourism managers, each ML model provides unique insights. For example, the RF model offers a powerful tool for developing reliable and accurate predictive models, particularly for tasks such as predicting traveler preferences, booking behaviors, or trip outcomes, thereby enabling more informed and data-driven strategies. LR models remain useful when prioritizing simplicity and interpretability, albeit with a slight trade-off in accuracy. Caution is advised when employing ANN models, especially with smaller tourism datasets, due to their higher risk of overfitting. Strategically combining these approaches can help balance accuracy, transparency, and complexity, ultimately enhancing decision-making in tourism marketing, customer segmentation, and service personalization.

Overall, these managerial implications underscore the need for travel businesses to adopt a strategic, data-driven approach to social media marketing in order to influence consumer behavior and drive engagement.

*Limitations & future work*

Although the current study explores a timely and relevant relationship foregrounding ASR in the travel planning behavior of tourists, it has its limitations. First, we only considered data from Facebook and Instagram. The study could be strengthened by investigating the content-sharing behavior of young users on several other SM platforms, including the hashtagging behavior, on Twitter and short video reels on platforms such as TikTok, Telegram, Snapchat, ShareCat, Moj, Josh, Koo, Clubhouse, Reddit, Quora, etc. Secondly, the data collected in the study shows that almost all the travel destinations for Indian tourists have been domestic. International trips may reveal interesting patterns among Indian tourists. One can investigate whether Indians prefer to visit developed countries like North America and Europe or explore popular neighborhood destinations like Malaysia, Thailand, Vietnam, Singapore, the United Arab Emirates, etc. Third, the current study has unpacked why specifically tourists chose a specific destination with an emphasis on SR. However, the potential investigation of tourism attitudes in terms of financial capabilities to avail leisure trips as a factor, as well as sustainable choices in travel behavior, has not been explored. A combination of financial constraints, sustainability motivations, and psychological traits of Gen Z, such as digital fatigue, along with ASR, can yield new insights. Finally, more sophisticated predictive modeling techniques, such as deep neural networks with a more extensive data set, may be investigated for more accurate and reliable predictions.

Future work can be undertaken to understand how shared experiences on SM platforms can help shape destination branding for destination marketing organizations (DMO) (Morosan, 2013), where DMO professionals would want to utilize participating tourists and their social network behavior to enhance the content of destination marketing websites. Smart tourism (Samancioglu, 2024) is another important discussion in the current tourism management literature. Leveraging tourists' self-generated content in conjunction with smart technologies (Balakrishnan et al., 2023) has the potential to enhance the development of more attractive and efficient tourist destinations. This approach can also contribute to a heightened sense of gratification among tourists (Liao et al., 2021) — particularly in their photographic and related behaviors—and promote an improved quality of life for local communities residing near destination sites (Shafiee et al., 2022). Future work can investigate the role of ASR in smart tourism technologies (Giaconne & Bonacini, 2019) to understand cultural heritage tourism and community engagement in India, since both ASR and the idea of smart tourism technologies incorporate digital participatory platforms.

Kutner, M.H., Nachtsheim, C.J., Neter, J. & Li, W. (2005). *Applied Linear Statistical Models*. 5th Edition, McGraw-Hill, Irwin, New York.

Lam, R., Cheung, C., & Chan, B. (2024). The mediating roles of travel motives and social capital on the relationship between cultural intelligence and general life satisfaction. *Asia Pacific Journal of Tourism Research*, *29*(2), 127–143.

Li, Y. W., & Wan, L. C. (2025). Inspiring tourists' imagination: How and when human presence in photographs enhances travel mental simulation and destination attractiveness. *Tourism Management, 106*, 104969.

Li, Y., & Xie, Y. (2020). Is a picture worth a thousand words? An empirical study of image content and social media engagement. *Journal of Marketing Research, 57*(1), 1-19.

Li, H., Zhang, L., & Hsu, C. H. (2023). Research on user-generated photos in tourism and hospitality: A systematic review and way forward. *Tourism Management*, *96*, 104714.

Liao, J., Wang, Y., Tsai, C., & Zhao, B. (2021). Gratifications of travel photo sharing (gtps) on social media: Scale development and cross-cultural validation. *Tourism Analysis, 26*(4), 265–277.

Liu, J., Wang, Y., & Chang, L. (2023). How do short videos influence users' tourism intention? A study of key factors. *Frontiers in Psychology, 13*, 1036570.

Luna-Cortés, G. (2021). Self-congruity, destination brand, and the use of social media. *Tourism Analysis, 26*(1), 77-81.

Mahapatra, S., Ray, S., & Mukherjee, S. (2025). Travel influencers' impact on followers' engagement behavior. *Tourism Recreation Research*, 1–18. https://doi.org/10.1080/02508281.2025.2484737

Maiberger, T., Schindler, D., & Koschate-Fischer, N. (2024). Let's face it: When and how facial emojis increase the persuasiveness of electronic word of mouth. *Journal of the Academy of Marketing Science*, *52*(1), 119–139.

McCabe, S., & Stokoe, E. (2010). Have you been away? Holiday talk in everyday interaction. *Annals of Tourism Research, 37(4)*, 1117-1140.
34